\documentclass[twocolumn,twoside,preprintnumbers,amsmath,amssymb,showkeys]{revtex4}
\usepackage{epsfig}
\usepackage{graphicx}% Include figure files

\usepackage{fancyhdr}

\usepackage{pslatex}

\begin{document}

%\title{Brazilian Journal of Physics instructions for the preparation of
%a 2-column format paper in \LaTeX}
\title{Source chaoticity in relativistic heavy ion collisions at SPS and
RHIC}

%\author{Fernando Silva$^1$, Paulo Machado$^1$ and Michael Schiff$^2$}
\author{Kenji Morita$^1$\footnote{Present address: Institute of
Physics and Applied Physics, Yonsei University, Seoul 120-749,
Korea. Electronic address: {\tt morita@phya.yonsei.ac.kr}}, 
Shin Muroya$^2$ and Hiroki Nakamura$^1$}

\affiliation{$^1$ Department of Physics, Waseda University, Tokyo
         169-8555, Japan \\
         $^2$ Matsumoto University, Matsumoto 390-1295, Japan}

%\received{on ????????????, 2006}

\begin{abstract}
 We investigate degree of coherence of pion sources produced in
 relativistic heavy ion collisions using multi-particle interferometry.
 In order to obtain ``true'' chaoticity, $\lambda^{\text{true}}$ from
 two-pion correlation
 functions measured in experiments, we make a correction for long-lived
 resonance decay contributions. Using this $\lambda^{\text{true}}$  and
 the weight factor which are obtained from parameter fitted to two- and
 three-pion correlation function, we calculate a chaotic fraction
 $\varepsilon$ and
 the number of coherent sources $\alpha$ for different colliding
 energies. The result gives constraints on the source and shows an
 increase of the minimum number of $\alpha$ with multiplicity, although
 multiplicity independent chaoticity is not excluded.


\end{abstract}
\maketitle

\thispagestyle{fancy}

\setcounter{page}{1}

\section{Introduction}

Relativistic heavy ion collisions provide us an unique opportunity to explore
nature of hot and dense nuclear matter on the earth. In the highest
energy collisions at the BNL-RHIC, it is expected that the matter
created soon after the collision of two heavy nuclei can be the strongly
interacting quark-gluon plasma, which gradually cools down and then
becomes hadronic matter via phase transitions. To understand the nature
of the QCD matter, it is important to know what kind of information
experimental observables contain. Pion interferometry has been one of the
most important observables because it can give us information on sizes
of the source which pions come from, through the HBT effect. The HBT
effect is a quantum-mechanical effect due to symmetrization of two-boson
wave function and occurs if the source is not completely coherent.
The strength of the two-boson momentum intensity correlation
takes its maximum value in the case of the perfectly chaotic source. 
Because the chaoticity can be related to the degree of thermal
equilibrium and to composition of the source, analyses of the chaoticity
can provide information on a state of hadronic matter which may reflect
how hadrons are produced.

Experimentally measured two-pion chaoticity
$\lambda=C_2(p,p)-1$ does not usually reflect real coherence of the source
because of long-lived resonance decay contributions. Three-pion
correlation function is more useful for this purpose. The three-pion
correlator measured in some experiments is 
\begin{align}
 r_3&(Q_3)= \nonumber\\
 &\frac{[C_3(Q_3)-1]-[C_2(Q_{12})-1]-[C_2(Q_{23})-1]-[C_2(Q_{31})-1]}{\sqrt{[C_2(Q_{12})-1][C_2(Q_{23})-1][C_2(Q_{31})-1]}},\label{eq:r3}
\end{align}
where $Q_{ij}=\sqrt{-(p_i-p_j)^2}$,
$Q_3=\sqrt{Q_{12}^2+Q_{23}^2+Q_{31}^2}$, and $C_n$ denotes the
$n$-particle correlation function. The weight factor $\omega=r_3(0)/2$
characterizes the degree to how pions sources are chaotic. For a completely
chaotic source, $\omega=1$. If the source chaoticity is really a
physical quantity, both two- and three-pion
correlation function should give quantitatively consistent value of the
chaoticity. However, experimental results do not seem so, because of the
apparent reduction of the
$\lambda$ due to long-lived resonances. But it has been shown that we
can impose stronger constraints by using both of two- and three-pion
correlation data \cite{Nakamura_PRC66}. 

In this work, we find that such decay contributions can be eliminated by
making use of a statistical model \cite{Morita_PTP114} and applied it to
various experimental data measured in SPS and RHIC experiments
\cite{Morita_PTP116}. Then, we obtain the weight factor $\omega$ from
two- and three-pion correlation function data in these experiments.
Using the resonance-corrected $\lambda$ which we call
$\lambda^{\text{true}}$ and $\omega$, we calculate a chaotic fraction
$\epsilon$ and the number of coherent sources based on a partially
multicoherent source model \cite{Nakamura_PRC61}. From this result, we
discuss how the structure of the sources changes from low energy
collisions at SPS and higher ones at RHIC \cite{Note1}.

\section{Correction of $\lambda$ factor for long-lived resonance decay}

In the presence of long-lived resonance decay contributions to two-pion
correlation function, the two-pion chaoticity for a chaotic source is
reduced as
\begin{equation}
 \sqrt{\lambda^{\text{eff}}}=1-\frac{N_\pi^{\text{r}}}{N_\pi}
\end{equation}
where $N_\pi$ is the total number of pions and $N_\pi^{\text{r}}$ is the
number of pions coming from long-lived resonance decay
\cite{Csorgo_Zphys71}. Hence, if one can estimate the ratio of pions
from the decay to the total number, one can obtain apparent reduction
factor caused by the decay contribution. Indeed, it is possible with the
help of the statistical model which has been known to give a nice
description of observed particle number ratio. Under the assumption of
the constant chemical freeze-out temperature and chemical potential and
boost-invariance along collision axis, the ratio of particle number can
be written as the ratio of number density, \textit{i.e.,}$N_i/N_j=n_i/n_j$,
where $n_i$ is given by
\begin{equation}
 n_i = \frac{g_i}{2\pi^2}\int_{0}^{\infty} dp p^2 f(E,T,\mu)
\end{equation}
with $f(E,T,\mu)$ being the equilibrium distribution functions
\cite{Cleymans_PRC60}. Using these formulae, we calculate the
$\lambda^{\text{eff}}$ for S+Pb and Pb+Pb collisions at the SPS and
Au+Au collisions at the RHIC and obtain $\lambda^{\text{true}}$ through
a relation $\lambda^{\text{true}}=\lambda^{\text{exp}}/\lambda^{\text{eff}}$
where $\lambda^{\text{exp}}$ is momentum-averaged experimental data
\cite{Morita_PTP114,Morita_PTP116}. The temperature and baryonic
chemical potential are determined by the $\chi^2$ fitting to
experimental data of particle ratio. See Ref.~\cite{Morita_PTP116} for
details. Here we summarize the result in Table~\ref{tbl:lambda}.
\begin{table}[htb]
 \begin{center}
  \caption{Summary of $\lambda^{\text{true}}$}\label{tbl:lambda}
  \begin{tabular}{cccc}\hline
   System &$(T,\mu_{\text{B}})$[MeV]&$\lambda^{\text{exp}}$&
   $\lambda^{\text{true}}$ \\ \hline
   S+Pb (NA44)&(173,196)&0.59(6)&0.94(6) \\
   Pb+Pb (NA44)&(161,223)&0.55(3)&0.98(3) \\
   Pb+Pb (WA98)&(161,223)&0.58(4)&1.03(4) \\
   Au+Au (STAR)&(158,36)&0.57(6)&0.93(8) \\ \hline
  \end{tabular}
 \end{center}
\end{table}

\section{Extraction of the weight factor}

In order to obtain the weight factor $\omega$, we have to extrapolate
experimentally measured $r_3(Q_3)$[Eq.~\eqref{eq:r3}] to $Q_3=0$.
Using a simple source function in which instantaneous emission and
spherically symmetric source are assumed, we construct the two- and the
three-pion correlation functions with the formulae
\begin{align}
  C_2(\boldsymbol{p_1,p_2})&=1+\lambda_{\text{inv}}\frac{f_{12}^2}{f_{11}f_{22}} 
  \label{eq:c2c},\\
 C_3(\boldsymbol{p_1,p_2,p_3}) &= 
 1+\nu\left(\sum_{(i,j)}\frac{f_{ij}^2}{f_{ii}f_{jj}}
 +2\frac{f_{12}f_{23}f_{31}}{f_{11}f_{22}f_{33}}\right) \label{eq:c3c},
\end{align}
where $f_{ij}=1/\sqrt{\cosh(R|{\bf q}_{ij}|)}$ is the source
function and $R$, $\lambda_{\text{inv}}$ and $\nu$ are
parameters which should be determined by a simultaneous $\chi^2$ fit to
the two- and the
three-pion correlation functions \footnote{Eq.~\eqref{eq:c3c} is a
simple reduction of the correlator part of the
fully chaotic case shown in Ref.~\cite{Nakamura_PRC61}. See
Ref.~\cite{Morita_PTP114} for discussion.}. From a set of
these parameters, we can
calculate $\omega$ using Eqs.~\eqref{eq:c2c},\eqref{eq:c3c} and
\eqref{eq:r3}. The results for $\omega$ are shown in Table~\ref{tbl:omega}.
\begin{table}[ht]
 \begin{center}
  \caption{Results for $\omega$}\label{tbl:omega}
  \begin{tabular}{ccccc}\hline
   &S+Pb(NA44)&Pb+Pb(NA44)&Pb+Pb(WA98)&Au+Au(STAR) \\\hline
   $\omega$&0.40$\pm$0.44&1.15$\pm$0.67&0.78$\pm$0.44&0.872$\pm$0.097 \\ \hline
  \end{tabular}
 \end{center}
\end{table}

\section{Analysis with the partially multicoherent model}
From considerations in previous sections, we have obtained the two 
quantities, $\lambda^{\text{true}}$ and $\omega$ as experimental
results. Next, we investigate the
coherence of the sources using the partially multicoherent model
\cite{Nakamura_PRC61}. In this model, there are two characteristic parameters
which are related to the $\lambda^{\text{true}}$ and $\omega$ as
\begin{align}
 \lambda^{\text{true}} &=
 \frac{\alpha}{\alpha
 +(1-\varepsilon)^2},\label{l-pm}
 \\
 \omega &= \frac{2\alpha^2 + 2\alpha
 (1-\varepsilon)^2 + 3(1-\varepsilon)^3
 (1-2\varepsilon)}
 {2[\alpha^2  + 3\alpha
 (1-\varepsilon)^2  +(1-\varepsilon)^3]} 
%\nonumber\\
%&\times
 \sqrt{\frac{\alpha+(1-\varepsilon)^2}{\alpha}}\label{ome-pm}.
\end{align}
By solving the above equations with respect to $\varepsilon$ and
$\alpha$ (this can be analytically), we can obtain allowed regions for
$\varepsilon$ and $\alpha$ corresponding to the available range of
$\lambda^{\text{true}}$ and $\omega$.

\begin{figure}[htb]
 \begin{center}
  \includegraphics[width=0.24\textwidth]{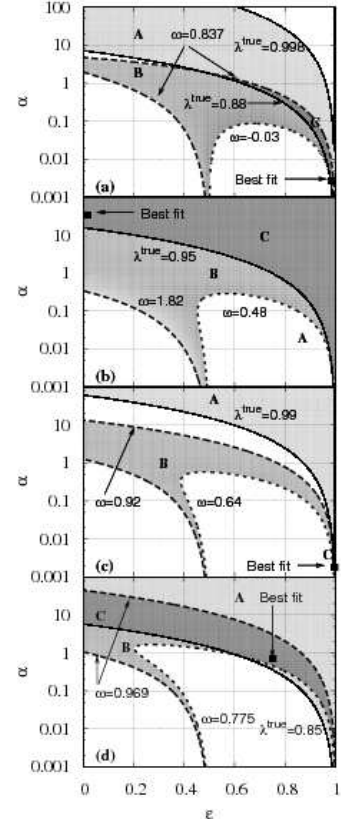}
  \caption{Allowed region for $\varepsilon$ and $\alpha$. From top to
  bottom, each figure (a)-(d) shows S+Pb, Pb+Pb by NA44, Pb+Pb by WA98
  and Au+Au case, respectively.}\label{fig:e-a}
 \end{center}
\end{figure}

The result is shown in Fig.~\ref{fig:e-a}. In each of figures, the
lightest shade area labelled ``A'' and the second one labelled ``B''
denote the allowed regions coming from $\lambda^{\text{true}}$ and
$\omega$, respectively. The darkest areas labelled ``C'' are overlap of
area ``A'' and ``B'', then correspond to the allowed parameter region
for $\varepsilon$ and $\alpha$. The best fit points are indicated by the
filled box.

From Fig.~\ref{fig:e-a}, it seems to be difficult to find systematic
change of the allowed regions. This result mainly comes from the fact
that $\lambda^{\text{true}}$'s are close to unity in the Pb+Pb
data. However, it has
been suggested that Coulomb correction to the two-pion correlation
functions is over-corrected one; the values of $\lambda^{\text{exp}}$ can
decrease if we take account of the partial Coulomb
correction \cite{CERES,PHENIX}. Hence, if appropriate corrections were
made, obtained $\lambda^{\text{true}}$ becomes smaller. 
In order to obtain a rough sketch of a tendency, we also draw the
allowed regions for $\lambda^{\text{true}}_{\text{pc}}$ which is 0.8
times smaller than $\lambda^{\text{true}}$. For S+Pb data, we multiply the
original $\lambda^{\text{true}}$ by factor 0.7 because Gamow (point-like
source) correction is made for this data.

\begin{figure}[htb]
 \begin{center}
  \includegraphics[width=0.24\textwidth]{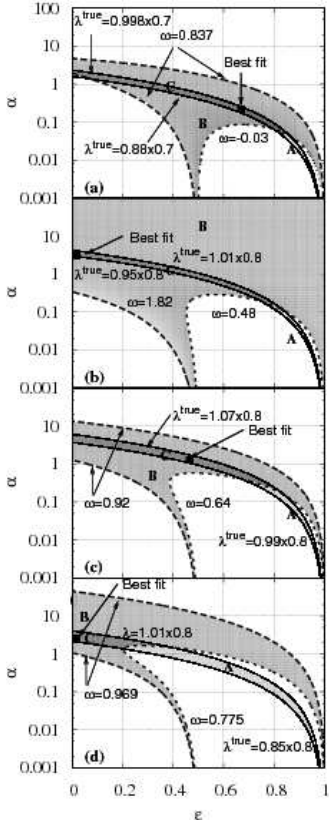}
  \caption{Allowed region for $\varepsilon$ and $\alpha$, in the case of using
  $\lambda^{\text{true}}_{\text{pc}}$. From top to
  bottom, each figure (a)-(d) shows S+Pb, Pb+Pb by NA44, Pb+Pb by WA98
  and Au+Au case, respectively.}\label{fig:e-alam}
 \end{center}
\end{figure}

\begin{figure}[ht]
 \begin{center}
  \includegraphics[width=0.32\textwidth]{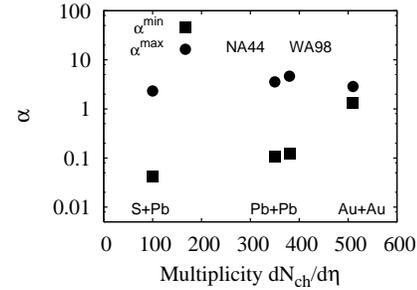}
  \caption{Maximum and minimum values of $\alpha$ as a function of
  multiplicity. For Pb+Pb data, both plots are slightly shifted along
  horizontal axis for a clear comparison of the results.}
  \label{fig:alpha}
 \end{center}
\end{figure}

From Fig.~\ref{fig:e-alam}, we can see that all allowed regions (the
darkest shaded areas) are narrow but there seem to exist systematics.
The best fit point seems to move upper left (small chaotic
fraction and large number of coherent sources) side, except for the NA44
Pb+Pb result. Most important feature of this result is that the upper limit
of $\varepsilon$ and lower limit of $\alpha$ are determined by the lower
limit of $\omega$. We plot the maximum and the minimum values of
$\alpha$ in Fig.~\ref{fig:alpha}. The clear increase of minimum number
of the coherent sources can be seen as a function of multiplicity while
maximum number of those shows no such clear tendency.

In summary, we have given an analysis of the degree how chaotic the pion
sources are in relativistic heavy ion collisions at the SPS and the
RHIC. The analysis can be done by using both two-pion correlations and
three-pion correlations. We find that the correction for long-lived
resonance decay contributions to the two-pion correlation function can
be subtracted with the help of the statistical model. 
From a point of view in which multicoherent sources and a background chaotic
source are produced, we show that the model gives constraints on the
structure of the source. Although the maximum number of the coherent
sources does not show a clear multiplicity dependence, the minimum
number of coherent source increases as the multiplicity increases. 
 
\noindent{\bf Acknowledgements}\\
The authors would like to thank Prof.~I.~Ohba and Prof.~H.~Nakazato for
their encourgement. K.~M's work is supported by a Grant for the 21st
Century COE Program at Waseda University from Ministry of Education,
Culture, Sports, Science and Technology of Japan and BK21 (Brain Korea
21) program of the Korean Ministry of Education.

%\smallskip 

\end{document}